\documentstyle[12pt,aasms4]{article}
\newcommand{\etal}{{\it et al.} }
\newcommand{\asca}{{\it ASCA} }

\begin{document}

\title{OCCULTATION MAPPING OF THE CENTRAL 
ENGINE IN THE ACTIVE GALAXY MCG $-$6$-$30$-$15}

\author{Barry McKernan\altaffilmark{1,2} and Tahir Yaqoob\altaffilmark{3} }

\altaffiltext{1}{Department of Physics and Astronomy, University of Leeds, 
Leeds LS2 9JT, UK}
\altaffiltext{2}{Present address: Department of Experimental Physics, University College Dublin, Dublin 4, Ireland}
\altaffiltext{3}{Laboratory for High Energy Astrophysics, Code 660.2
Goddard Space Flight Center, Greenbelt, MD 20771}

\accepted{23 April 1998}
\vspace{6cm} 
\centerline{\it Accepted for publication in the Astrophysical Journal Letters}

\begin{abstract}
The colossal power output of active galactic nuclei (AGN)
is believed to be fueled by the accretion of matter 
onto a supermassive black hole. 
This central accreting region of AGN has hitherto been 
spatially unresolved and its structure therefore unknown. 
Here we propose that a previously reported
`deep minimum' in the X-ray intensity of the AGN MCG$-$6$-$30$-$15,
was due to a unique X-ray occultation event and that 
it probes structure of the central
engine on scales $<  10^{14} \ \rm cm$, or 
$1.4\times 10^{-7}$ arcseconds. 
This resolution
is more than a factor of $\sim 3\times 10^{6}$ greater than is possible
with current X-ray optics.
The data are consistent with a bright central source surrounded
by a less intense ring,
which we identify with the inner edge of an  
accretion disk. These may be the first direct measurements of the spatial
structure and geometry 
of the accreting black-hole system in an active galaxy.
We estimate a mass lower limit for sub-Eddington accretion of
$3.1\times 10^{5} M_{\odot}$. If the ring of
X-ray emission is identified with the inner edge of
an accretion disk, we get mass upper limits of 
$1.9\times10^{8}$ and 
$9.1\times10^{8} M_{\odot}$ for a
non-rotating and 
maximally rotating black hole respectively.
We point out that our occultation interpretation is controversial in
the sense that X-ray variability in AGNs is normally attributed to
intrinsic physical changes in the X-ray emission region, such as
disk or coronal instabilities. 
\end{abstract}


\section{Introduction}
\label{intro}

The accretion of matter onto a supermassive black hole as a
mechanism for fueling the output of active galactic nuclei
(AGN) is a paradigm 
strongly supported by recent spectroscopic observations  
of the iron K$\alpha$ X-ray emission 
line (Tanaka \etal 1995; Yaqoob \etal 1995; Nandra \etal 1997a and
references therein). 
The extreme
Doppler and gravitational energy shifts of the line photons, together
with the shape of the line are consistent with an origin in a disk 
rotating about a black hole (Fabian \etal 1989; Laor 1991).
Strong gravitational redshifts, in which photon energies are
changed by more than 10\%, occur
only when matter approaches closer than $\sim  20$ gravitational radii
($= 20r_{g}; r_{g} \equiv GM/c^{2}$)
from a compact object.
However, it is not possible to directly map the physical structure
of the system since the highest spatial resolution of 
X-ray optics
technology is a factor $\sim 10^{6}$ too poor for even 
the closest AGN.  
Optical and radio observations provide greater resolution but the
bulk of the emission at these wavelengths is not generated close
enough to the central engine.
So far, the highest spatial resolution observations,
at radio wavelengths, have revealed
a Keplerian disk in the AGN NGC 4258 down to only $\sim 60,000 r_{g}$ 
(Miyoshi \etal 1995; Maoz 1995).
This still falls short by a factor $\sim 3000$ of mapping the 
black-hole region.

The AGN MCG$-$6$-$30$-$15 ($z= 0.008$) was observed by the X-ray 
astronomy satellite {\it ASCA}
({\it Advanced Satellite for Astrophysics and Cosmology};
Tanaka, Inoue, and Holt 1994) for $\sim 4.2$ days on 23 July 1994.
Results from this observation 
have already appeared in the literature,
including the  
0.5--10 keV lightcurve
(Iwasawa \etal 1996, hereafter I96; Reynolds 1997; Yaqoob \etal 1997; see
Figure 1) and a broad, asymmetric, variable iron K line with
 a strong red wing,  consistent with a disk inclined at 
$\sim 30^{\circ}$ rotating about a black hole
(Tanaka \etal 1995; I96).
The X-ray luminosity exhibits erratic
variability
on all timescales down to $<50$ s 
(Matsuoka \etal 1990; Green, McHardy, and Lehto 1995; Reynolds \etal 1995;
Nandra \etal 1997b).
Causality arguments alone cannot put constraints on the size of
the X-ray emission region since the high-frequency, lower amplitude
variability may occur at localized
regions of the source.
Figure 1 shows an extended intensity dip at the end of the observation, 
from $\sim 3.3\times10^{5}$s to $\sim 3.6\times10^{5}$s. This
feature has previously been dubbed as the 'deep minimum', or DM
(I96).
A closer inspection
(Figure 2a) reveals a remarkable (albeit approximate) symmetry about
the minimum luminosity.
 
We propose that the dip was caused by an occultation of the X-ray source by 
optically-thick matter. 
This interpretation is controversial and 'non-standard', as X-ray
variability in AGNs is normally attributed to intrinsic properties
of the X-ray emission region, such as disk or coronal instabilities.
However, so little is know about structure of the central engine in
AGNs that the occultation scenario should be explored.
The obscurer must be optically thick
because the dip continuum spectrum 
only shows evidence for {\it weak} absorption, nowhere near enough
to explain the observed intensity variation over the whole \asca
bandpass (see Weaver \& Yaqoob 1998, hereafter WY98).
The luminosity at the absolute minimum of the dip is $\sim 0.4$ of 
the pre-dip value and must represent persistent
emission which has much smaller surface brightness than the primary source.
Hereafter we will refer only to the primary X-ray emission, unless
explicitly referring to the persistent emission.
The proposed obscurer very likely hides the most compact and variable
part of the X-ray source, since 
the usual rapid variability outside the dip 
is absent during the obscuration.

Of course, it is possible that the dip is due to intrinsic variation of the
X-ray source. However, the origin of X-ray variability in AGN is
not understood. Models 
which come close to successfully reproducing 
the observable quantities obtained from
AGN lightcurves are of the shot-noise or `rotating hot-spot'
variety (e.g. Green \etal 1993; Bao and Abramowicz 1996 and references
therein).
However, the
parameters of such models must
be highly tuned in order to reproduce AGN power spectra.
On the other hand, we show in this paper that a very simple-minded 
model can account for the temporal profile of the intensity dip
in MCG $-$6$-$30$-$15 and briefly discuss the implications for 
AGN X-ray variability in general.

\section{A Simple Model for the Dip Temporal Profile}

Here we present a simple occultation model of the intensity
dip (Figure 2).
The dip is clearly divisible into several 
distinct time intervals. The turning points are labelled
$t_{1}$--$t_{8}$ in Figure 2 and
the values shown for $t_{2}$--$t_{7}$
are the nearest centers
of the 512 s bins. The values for 
$t_{1}$ and $t_{8}$ are obtained
from a model described below. Without any further analysis, one
can make some immediate deductions, since the obscurer cannot travel
faster than the speed of light. The duration of the observed
dip ($\sim 3 \times 10^{4}$ s) 
implies a source size $< 9\times 10^{14} \ \rm cm$ and
the duration of the absolute
minimum, $\sim 3000$ s, implies that distance scales 
$< 9 \times10^{13} \ \rm cm$ are resolved.  

The fact that there are two ingresses and egresses in the dip implies that
either the obscurer or the X-ray emission must be spatially non-uniform.
A non-uniform obscurer must
have an extremely contrived shape 
and size relative to the X-ray source. 
For a non-uniform source, 
we can deduce a one-dimensional intensity profile. This profile
must peak in three places to produce
the nearly symmetric occultation profile observed. Two of the peaks
 must have similar intensity and
be approximately equidistant from the central peak to preserve
approximate symmetry.
The obscurer must be larger than the distance spanned by the three
peaks to avoid local maxima in the occultation profile. 
As the obscurer moves over the three peaks, if the first and last peaks
are the more intense, then the size of the obscurer must be
fine-tuned so that the last peak is covered at the same
rate as the first peak is uncovered, in order to avoid a 
large, sharp rise after the absolute minimum. Moreover, the
duration of the absolute minimum must be significantly greater than
the duration of the inflexions in the dip ($t_{2}$--$t_{3}$ and 
$t_{6}$--$t_{7}$), which is not observed.  
Therefore, the central peak must have the greatest intensity. 
Thus, the 
one-dimensional intensity
profile must have the general (but not necessarily exact) form shown 
in Figure 2b.
In this case, the first drop in intensity ($t<t_{1}$) is small
and not observable since most of the X-ray emission is still unobscured.
The large drop over $t_{1}$--$t_{2}$ is then due to the central
source being covered. 
Note that the high data point when the source has just come out
of the dip is {\it not} required to be explained by 
the model.
This is because the central source, which
we know exhibits rapid and erratic variability, is then completely uncovered.
 
Although any two-dimensional model which gives the profile in Figure 2b 
is valid, it appears that an 
 obvious and the least contrived 
 realization of the inferred intensity profile is 
an X-ray
source consisting of a high-luminosity central part, surrounded by
a ring of dimmer emission. The gap between the central source and the
ring is {\it required}, in order
to fit the intensity profile between $t_{2}$--$t_{3}$
and $t_{6}$--$t_{7}$.  We construct a simple
two-dimensional model which can explain
the data (Figure 2b). The central source is represented
by a circle of radius $r_{s}$, and the obscurer  by a circle of radius
$r_{0}$. In reality the latter may be 
more elongated in the direction of travel,
 but the important quantity is its
length, $2r_{0}$. The inner and outer ring radii are $r_{1}$ and
$r_{2}$ respectively. To account for
the slight asymmetry in the dip
(luminosity during $t_{6}$--$t_{7}$ 
higher than $t_{2}$--$t_{3}$), the half-ring {\it uncovered last} is made
 more luminous.
 Let the intensities per unit area of the
central source, first and last {\it covered} halves 
of the disk be $I_{s}$, $I_{b}$, $I_{r}$ respectively and let $v$ be the 
transverse velocity of the obscurer. We find a fiducial model of the 
dip profile by adjusting $r_{s}/v$, 
$I_{s}/I_{b}$ and $I_{s}/I_{r}$. 
The other parameters are fixed 
by the relations: $(r_{1}-r_{s})/v= t_{3}-t_{2}$, 
$(r_{2}-r_{1})/v=t_{4}-t_{3}$ 
and $2(r_{0}-r_{2})/v= t_{5}-t_{4}$.
A good fit 
(solid line in Figure 2a) is obtained with $r_{s}/v = 1536$s,
 $r_{1}/r_{s} =3.66$,
$r_{2}/r_{s}=7.32$, $r_{0}/r_{s}=8.32$,
$I_{b}/I_{s}=.024$ and
 $I_{r}/I_{s}=.012$. 
The dip slope
between $t_{1}$--$t_{2}$ strongly
constrains the uncertainty in $r_{s}/v$ to $<20\%$ of
the fiducial value. The 
ranges on $r_{1}$, $r_{2}$ and $r_{0}$ depend on the 
uncertainties in $r_{s}$ and the turning points,
and again are $<20\%$, assuming
a tolerance of 512 s (i.e. the bin width) for the 
turning points.
The tolerances on $I_{b}/I_{s}$ and $I_{r}/I_{s}$
are also less than $20\%$.

One interpretation is to identify the emission ring with the inner
edge of an accretion disk.
An inclined disk would explain the asymmetry in the occultation since
Doppler effects due to rotation would give intensity enhancement or
reduction from blueshifts  or redshifts respectively. 
$I_{b}/I_{r} \sim 2$ can easily be obtained
for velocities of only $0.17c \ $ (Reynolds and Fabian 1997).
In practice one must integrate Keplerian velocities over the disk
(note, at $6r_{g}$, $v=0.4\ \rm c$).
The persistent continuum emission during the dip is likely to be
due to any uncovered part of the accretion ring plus
emission from the rest of the accretion disk (which must have much
lower surface emissivity than the inner ring).

\section{The X-ray Continuum and Fe-K Emission Line During the Dip}

We have shown that a simple-minded occultation model successfully
reproduces the observed intensity dip in MCG $-$6$-$30$-$15.
WY98 showed that the same model successfully explains the shape of the
Fe-K line during the dip. The peculiar shape during the dip
(bloated red wing and diminished blue-side emission) has
been interpreted by a number of authors as evidence for line-emission
from within $6r_{g}$ of black hole (I96; Reynolds and Begelman 1997;
Dabrowski \etal 1997). In the occultation model, the shape is due
to obscuration of the most blueshifted and redshifted part of the
line emission from the putative accretion disk, leaving only redshifted
emission. Line emission from within $6r_{g}$ is 
not {\it required}. Moreover, WY98 find, in contrast to I96, that  
a huge Fe-K line equivalent width (EW $ > 1000$ eV) 
during the dip is {\it not} required but is of the same order
as the EW from the 4.2-day time-averaged value ($\sim 400$ eV).
Further, the Fe-K line intensity is {\it not} required to increase
during the dip, compared to the time-averaged
or flaring-state value (in contrast to the I96 analysis). In the analysis
of WY98 the data are consistent with the line intensity {\it decreasing}
during the dip.

WY98 also re-analysed the X-ray continuum spectrum during the dip.
I96 interpreted the change in the 3--10 keV continuum power-law
photon index, $\Gamma$,
from $\sim 2$ (time-averaged `normal' value) to $\sim 1.7$ (dip value),
as an intrinsic continuum change. WY98 found that the dip data are 
consistent  with  $\Gamma$ fixed at the `normal' value of 1.98,
but allowing for extra absorption (equivalent neutral Hydrogen column
density of $\sim 2 \times 10^{22} \rm \ cm^{-2}$). WY98 interpreted
this as the increase in opacity of the photo-ionized absorber in
MCG $-$6$-$30$-$15 (e.g. Reynolds \etal 1995; Otani \etal 1996)
due to the central ionizing source being blocked by the putative
optically-thick obscurer. 
Thus the occultation model for the dip temporal profile is
entirely consistent with the observed Fe-K line and X-ray continuum
during the dip.

\section{Discussion}

If accreting matter is exposed to the same UV/X-ray
luminosity that we observe ($L \sim 4 \times 10^{43} \ \rm ergs
\ s^{-1}$) then for gravitational infall to overcome
outward radiation pressure requires the mass of the central
black hole, $M_{\rm BH}$, to exceed 
$3 \times 10^{5} M_{\odot}$.
Thus, if $r_{1}$ is the radius of 
the inner edge of the accretion disk,
identified with the last stable orbit of matter, and
$\Delta t$ is the time taken for the obscurer to traverse 
$r_{1}$ (i.e. $t_{3}-0.5[t_{2}+t_{1}$]),
then $\kappa r_{g}=r_{1}$
where $\kappa=6$ or 1.24 for Schwarzschild or 
maximally rotating Kerr metrics respectively.
But, $ r_{1}<c \Delta t$ and $r_{g} = 1.48 \times 10^{13} \ \rm 
(M_{\rm BH}/M_{\odot})$ cm, so
$M_{\rm BH} < 1.9 \times 10^{8} M_{\odot}$, or 
$<9.1\times 10^{8} M_{\odot}$,
 for a Schwarzschild 
or Kerr
black hole respectively. 
Assuming instead, the Keplerian velocity 
at the inner disk edge as the maximum 
velocity of the obscurer ($v/c = \sqrt{r_{g}/r} = \sqrt{1/\kappa}$)
yields smaller mass upper limits of
$M_{\rm BH}<7.7\times10^{7} M_{\odot}$
and
$M_{\rm BH} <8.2\times10^{8} M_{\odot}$ 
for a Schwarzschild and extremal Kerr metric respectively.
If the obscurer is at a 
distance, $d$, from
the central source, $r_{1} = v \Delta t = c \sqrt{r_{g}/d}$
gives
$d = c^{2}[\Delta t]^{2}/\kappa^{2}r_{g} = 
6 \times 10^{15}[\Delta t]^{2}/[\kappa^{2}
(M_{\rm BH}/M_{\odot})]$ cm.
Thus, $M_{\rm BH} > 3 \times 10^{5} M_{\odot}$
implies $d<1.8\times 10^{16}\ \rm cm$ and $d<4.2\times10^{17}\ \rm cm$ for
Schwarzschild and Kerr metrics respectively.
The origin of the optically-thick blobs is unspecified but 
Guilbert and Rees (1988) presented some simple arguments for the
existence of dense ($n > 10^{15} \rm \ cm^{-3}$),
optically thick matter residing at the heart of accreting
sources. The only independent estimate of the size of the `blobs'
is that they should be much thicker than $10^{9}/n_{15} \ \rm cm$
($n_{15}$ in units of $10^{15} \rm \ cm^{-3}$).     
The blobs must 
be optically thick even near their physical boundaries
(i.e they must have fairly
sharp edges), otherwise the dip profile would not be so well defined. 
Also, the blobs must be fairly stable, esepcially if they are created
in the central region itself and 'propelled' up to high altitudes.

The nature of the bright central X-ray source is intriguing.
An inclined jet is unlikely, since even at
$30^{\circ}$ the inflexions in the dip profile 
($t_{2}$--$t_{3}$ and $t_{6}$-$t_{7}$) would have
different durations.
If the central X-ray 
source extracts its energy directly from the black hole
then the metric is likely to be Kerr since energy cannot
be extracted from a non-rotating black hole 
(Blandford and Znajek 1977; see also Ghosh and Abramowicz 1997).
Occultations such as the one described here may occur 
frequently in AGN, but the relative sizes of the obscurer and
source must be just right in order to observe such a clear event.
Indeed the usual rapid variability or flicker, may be partly caused 
by the transit
of optically thick bodies smaller than the source.
We tested this hypothesis, again using a simple-minded model.
Representing the bright central source as a circular disk with
uniform emissivity (ignoring the ring due to its weaker emission),
optically-thick blobs with ranges in radii (relative to
the source) and velocities taken
from Gaussian distributions were passed over the source.
An additional parameter is required to specify the 'blob birth-rate'
(i.e. the rate at which new blob trajectories are started).
Such a model was used to produce predicted lightcurves for
different model parameters. The power spectrum of each lightcurve
was computed using the method of Papadakis and Lawerence (1993),
omitting Poisson noise. In the range $\sim 10^{-5}$ Hz to $\sim 10^{-2}$
Hz, no preferred or 'universal' power-law spectral slope was found.
It is possible to produce power-law spectra with slopes 
similar to those typically measured ($\sim -1$ to
$-2$) but for most parameter values, the slopes are too steep. 
Thus, fine-tuning would be necessary to explain the
`universal' power-law slopes found in the handful of AGN in
which it can be measured (e.g. Lawerence and Papadakis 1993; Green \etal
1993). This is essentially because if the 
product of blob crossing-time and birth-rate
is too large or too small, there will be no variability.
The direct simulated lightcurves assume, of course 
that the {\it intrinsic} source intensity is constant, which almost
certainly is not the case. Thus our model does not explain
AGN variability in general but its effects are potentially
important to consider in any model of AGN variability.

We thank the \asca TEAM and mission operations at ISAS, Japan,
for their efforts and hard work; Kim Weaver for her
work on the dip spectrum, and Karen Leighly, Paul Nandra for some
useful discussions. We also thank the anonymous referee.
This research made
use of archival data at the HEASARC, Laboratory for
High Energy Astrophysics, NASA/Goddard Space Flight Center.

\newpage

\section*{ Figure Captions}

\noindent
{\bf Figure 1}  0.5--10 keV X-ray lightcurve
from MCG$-$6$-$30$-$15 obtained by one of the CCD detectors
(known as SIS0) aboard {\it ASCA} (Tanaka \etal 1994).
The data are binned at 512 s. The reference time is 23 July 1994 
UT 05:56:13.

\noindent
{\bf Figure 2} 
(a) Close-up of the latter portion of the lightcurve in Figure 1.
Dashed line is the simultaneous background 
(not subtracted from the on-source data). The attitude of the 
satellite was stable to better then 0.004$^{\circ}$ and the 
apparent occultation event
is consistent in all four \asca detectors.
Solid line is the predicted occultation profile of a simple
two-dimensional model consisting of central source (radius
$r_{s}$), and ring with inner and outer radii, $r_{1}$ and
$r_{2}$ respectively. The intensities per unit area of 
the central source, first-covered and last-covered halves
of the disk are $I_{s}$, $I_{b}$, and $I_{r}$ respectively.
The obscuring body has a radius $r_{0}$. However it need not
be circular, the important quantity is the length in the direction
of travel. The solid line corresponds to a model with $I_{s}/I_{b}=
0.024$ and $I_{s}/I_{r} = 0.012$, $r_{s}/v=1536$ s, where 
$v$ is the transverse velocity of the obscurer.
The other parameters,
$r_{1}/r_{s}=3.66$, $r_{2}/r_{s}=7.32$, and $r_{0}/r_{s}=8.32$,
are determined by the turning points, $t_{2}$--$t_{7}$ which 
were estimated from the data to the nearest
512 s bins. The time intervals shown correspond to the intervals
between the front-end of the obscurer moving
over the regions marked by the corresponding dotted lines
(see text).
(b) Schematic of the source geometry (top)
and one-dimensional 
intensity profile (bottom) of the ring system,
summed in a direction perpendicular to the velocity of the obscurer.
In this model, the center of the obscurer passes over the center of
the source but this need not be the case. 
Note that the high data point when the source has just come out
of the dip is {\it not} required to be explained by
the model.
This is because the central source, which
we know exhibits rapid and erratic variability, is then completely uncovered.


\begin{references}

\reference{Baoo1996} Bao, G., \& Abramowicz, M. A. 1996, MNRAS, 465, 646

\reference{Blan1977} Blandford, R. D. \&\ Znajek, P. L. 1997, MNRAS
,197, 433

\reference{Dabr1997} Dabrowski, Y., Fabian, A. C., Iwasawa, K.,
Lasenby, A. N., \& Reynolds, C. S. 1997, MNRAS, 288, L11

\reference{Iwas1996} Iwasawa, K., \etal 1996, MNRAS, 282, 1038 (I96)

\reference{Fabi1989} Fabian, A. C., Rees, M. J., Stella, L.,
\& White, N. E. 1989, MNRAS, 238, 729

\reference{Ghos1997} Ghosh, P., \& Abramowicz, M. A. 1997, MNRAS, 292, 887

\reference{Gree1993} Green, A. R., McHardy, I. M. \&\ Lehto, H. J.  1993, MNRAS, 265, 664

\reference{Guil1998} Guilbert, P. W., \& Rees, M. J. 1988, MNRAS, 233, 475

\reference{Laor1991} Laor, A. 1991, ApJ, 376, 90

\reference{Lawr1993} Lawrence, A., \& Papadakis, I. E. 1993, ApJ, 414, L85

\reference{Maoz1995} Maoz, E.  1995, ApJ, 447, L91

\reference{Mats1990} Matsuoka, M., Piro, L., Makoto, Y. \&\ Murakami, T. 1990, ApJ, 361, 440

\reference{Miyo1995} Miyoshi, M. {\it et al.} 1995, Nat, 373, 127

\reference{Nand997b} Nandra, K. George, I. M., Mushotzky, R. F., Turner, T. J. \&\
Yaqoob, T. 1997b, ApJ, 476, 70

\reference{Nand997a} Nandra, K., George, I. M., Mushotzky, R. F., Turner, T. J. \&\
Yaqoob, T. 1997a, ApJ, 477, 602

\reference{Otani1996} Otani, \etal 1996, PASJ, 48, 211

\reference{Papa1993} Papadakis, I. E., Lawrence, A. 1993, MNRAS, 261, 612

\reference{Reyn1997} Reynolds, C. S. 1997, MNRAS, 286, 513

\reference{Rybg1997} Reynolds, C. S., \& Begelman, M. C. 1997, ApJ, 488, 109

\reference{Refa1997} Reynolds, C. S., \& Fabian,  A. C. 1997, MNRAS, 290, L1

\reference{Reyn1995}  Reynolds, C. S., \& Fabian,  A. C., Nandra, K.,
\& Inoue, H., 
Kunieda, H. \&\ Iwasawa, K. 1995, MNRAS, 277, 901

\reference{Tana1994} Tanaka, Y., Inoue, H., \& Holt, S. S. 1994, PASJ, 46, L37

\reference{Tana1995} Tanaka, Y., {\it et al.}, Nat, 375, 659

\reference{Weav1998} Weaver, K. A., \& Yaqoob, T. 1998, ApJ, submitted (WY98)
%(available by anonymous ftp from {\tt lheaftp.gsfc.nasa.gov} in
%{\tt pub/yaqoob/papers/mcg6\_fek/})

\reference{Yaqo1995} Yaqoob, T., Edelson, R., Weaver, K. A., Warwick, R. S.,
Mushotzky, R. F., Serlemitsos, P. J., \& Holt, S. S. 1995, ApJ,
453, L81 

\reference{Yaqo1996} Yaqoob, T., Serlemitsos, P. J., Turner, T. J., George, I.
M.,
\& Nandra, K. 1996, ApJ, 470, L27

\reference{Yaqo1997} Yaqoob, T., McKernan, B., Ptak, A., Nandra, K.,
Serlemitsos, P. J. 1997, ApJ, 490, L25

\end{references}
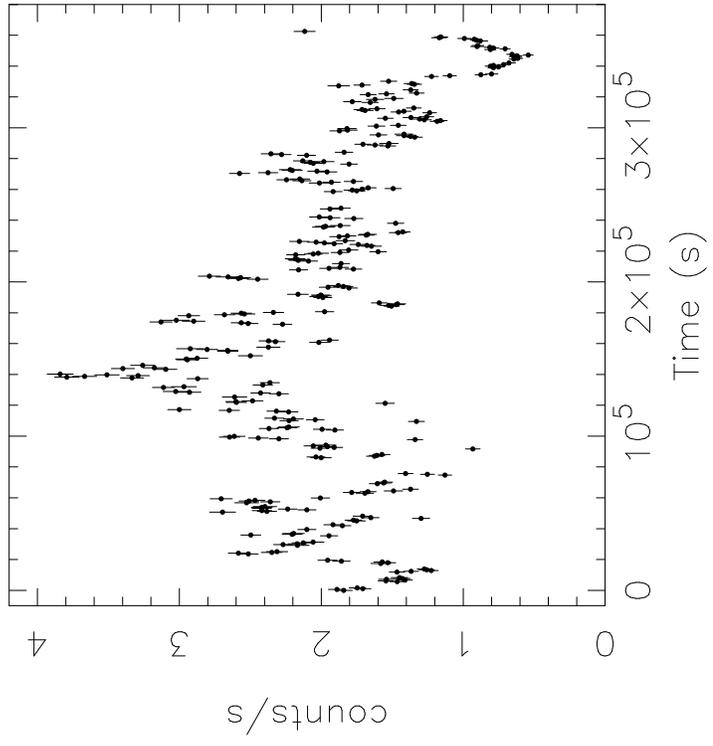
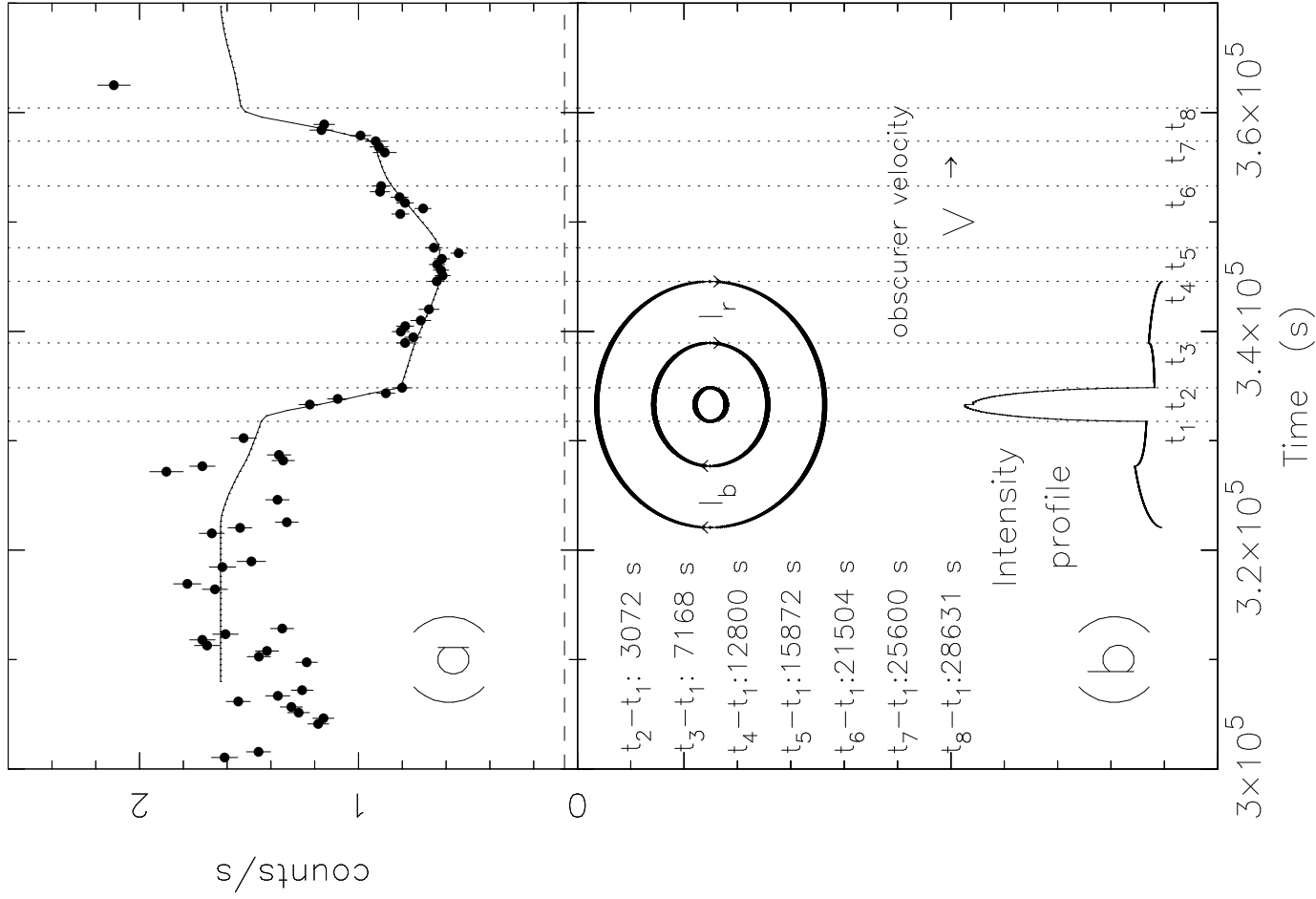
\end{document}